\documentclass[intlimits,twoside,a4paper]{article}
\usepackage{graphicx}
\usepackage{amsmath,amssymb}
\usepackage{wrapfig}

\usepackage[T2A]{fontenc}
\usepackage[cp1251]{inputenc}

\usepackage[eqsecnum]{cmpj2}

\def\be{\begin{equation}}
\def\ee{\end{equation}}
\def\bea{\begin{eqnarray}}
\def\eea{\end{eqnarray}}

\issue{2015}{18}{1}{13501}
\doinumber{10.5488/CMP.18.13501}

\title[]%
{Virial expansions and augmented van der Waals approach:
Application to Lennard-Jones-like Yukawa fluid
}
\author[A. Trokhymchuk, R. Melnyk, I. Nezbeda]{A. Trokhymchuk\refaddr{label1,label4}, R. Melnyk\refaddr{label1}, I. Nezbeda\refaddr{label2,label3}}
\addresses{
\addr{label1}
Department of the Theory of Solutions,
Institute for Condensed Matter Physics of the
National  Academy of Sciences of Ukraine,
 79011 Lviv, Ukraine
\addr{label4}
Institute of Applied Mathematics and Fundamental Sciences,
Lviv Polytechnic National University,  79013 Lviv, Ukraine
\addr{label2}
Faculty of Science, J.E. Purkinje University,
    400 96 \'{U}st\'{\i} nad Labem, Czech Republic
\addr{label3}
  E. H\'ala Laboratory of Thermodynamics, Institute of Chemical Process
   Fundamentals, Academy of Sciences,
    165 02 Prague 6, Czech Republic
}

\date{Received March 1, 2015}

\begin{document}

\maketitle

\begin{abstract}

We argue that recently  proposed [Melnyk et al., {Fluid Phase Equilibr.}, 2009, \textbf{279}, 1] a criterion to split the pair interaction energy into two parts, one of which is forced to be responsible the most accurate as possible for excluded volume energy in the system,
results in expressions for the virial coefficients that improve the performance of the virial equation of state in general, and at subcritical temperatures, in particular.
As an example, application to the Lennard-Jones-like hard-core attractive Yukawa fluid is discussed.

\keywords excluded volume, hard-core attractive Yukawa fluid, virial equation of state, second virial coefficient, van der Waals equation
\pacs  51.30.+i, 64.30.+t
\end{abstract}


\section{Introduction}
\label{sec:1}

Despite the significant progress in the development of modern tools in the statistical theory of
liquids~\cite{HM2013,Zwanzig,WidomSCI1963,BH1976,wcaSCI1980,HendersonMP1978,Shukla2000,tang2000,blum,duh},
there still are numerous studies where thermodynamic properties are expressed in the terms of virial expansions
(e.g., see references~\cite{NezbedaSmith,Masters2008,NareshSingh2009} and references therein).
The classical example of the virial expansion approach is the virial equation of state (EOS) \cite{Kamerlingh1901,MayerMayer1977}
\begin{equation}
\frac{p}{k_{\rm B}T} = \rho + B_2(T)\rho^2 + B_3(T)\rho^3 + B_4(T)\rho^4 + \ldots\,,
\label{Pvir}
\end{equation}
where $p$ is the pressure, $k_{\rm B}$ is the Boltzmann constant, $T$ is the temperature, $\rho$ is the number density and $B_n(T)$, $n=2, 3, 4, \ldots$ are the virial coefficients.
The success of the virial expansion approach relies, first of all, on the knowledge of the virial coefficients.
The first two virial coefficients, $B_2(T)$ and  $B_3(T)$,  by various techniques can be obtained experimentally, while in theoretical studies both can be relatively easily evaluated numerically (for some fluid models even analytically~\cite{FourthVirCoeff}) following their definition in terms of Mayer cluster integrals~\cite{MayerMayer1977}
\begin{equation}
B_2(T) = -\frac{1}{2V}\int\int f(r_{12})\rd{\bf r}_1\rd{\bf r}_2
\label{B2}
\end{equation}
and

\begin{eqnarray}
B_3(T) =
 -\frac{1}{3V}\int\int\int f(r_{12})f(r_{13})f(r_{23})\rd{\bf r}_1\rd{\bf r}_2\rd{\bf r}_3\,, 
\label{B3}
\end{eqnarray}

\noindent
where
\begin{equation}
f(r) = \exp[-\beta u(r)] - 1
\label{fM}
\end{equation}
is the Mayer function
and $u(r)$ is the pair interaction energy in a target model fluid.
The expressions for higher order virial coefficients are much more complicated, especially due to a significant increase in the number of distinct integrals that are required to be evaluated \cite{LabKolMal2005}.
Therefore, there are not so many potential functions $u(r)$, for which the virial coefficients $B_n(T)$ of the order $n>2$
are known.

Traditionally, the virial expansion approach is the most advanced for the model fluid composed of purely repulsive hard spheres of diameter $\sigma$. In this case, the virial coefficients $B_n$ are independent of temperature and have been calculated
up to the twelfth order~\cite{ClisbyMcCoy2006,Wheatley2013,ZhangPettitt2014,SchultzKofke2014}. Being explored in the virial EOS, these coefficients lead to the pressure $p^{\rm hs}$ of the hard-sphere (hs) fluid that is rather accurate in comparison with computer simulations data  for densities up to the fluid--solid transition~\cite{Woodcock2010}.

However, the success of the virial expansion approach is not so evident when apart from the hard-sphere repulsion,  the interaction potential $u(r)$ and, consequently, the exponential of the Mayer function in equations~(\ref{B2})--(\ref{fM}) both include the attractive interaction energy between molecules.
In this case, the virial expansions approach tends to diverge when approaching the thermodynamic states associated with condensation.
This fact imposes serious limitations on the applicability of the virial expansion approach to properly describe the vapour--liquid equilibrium in fluid systems.
The problem of virial expansion divergence, in the region of condensation, turned out to be a long-standing issue in the case of Lennard-Jones (LJ) fluid (e.g., see 
recent papers by Ushcats \cite{UshcatsPRL2012,UshcatsPRE2013,UshcatsJCP2013,UshcatsJCP2014} and references therein).

\begin{figure}[!t]
\centerline{
\includegraphics[width=0.5\textwidth]{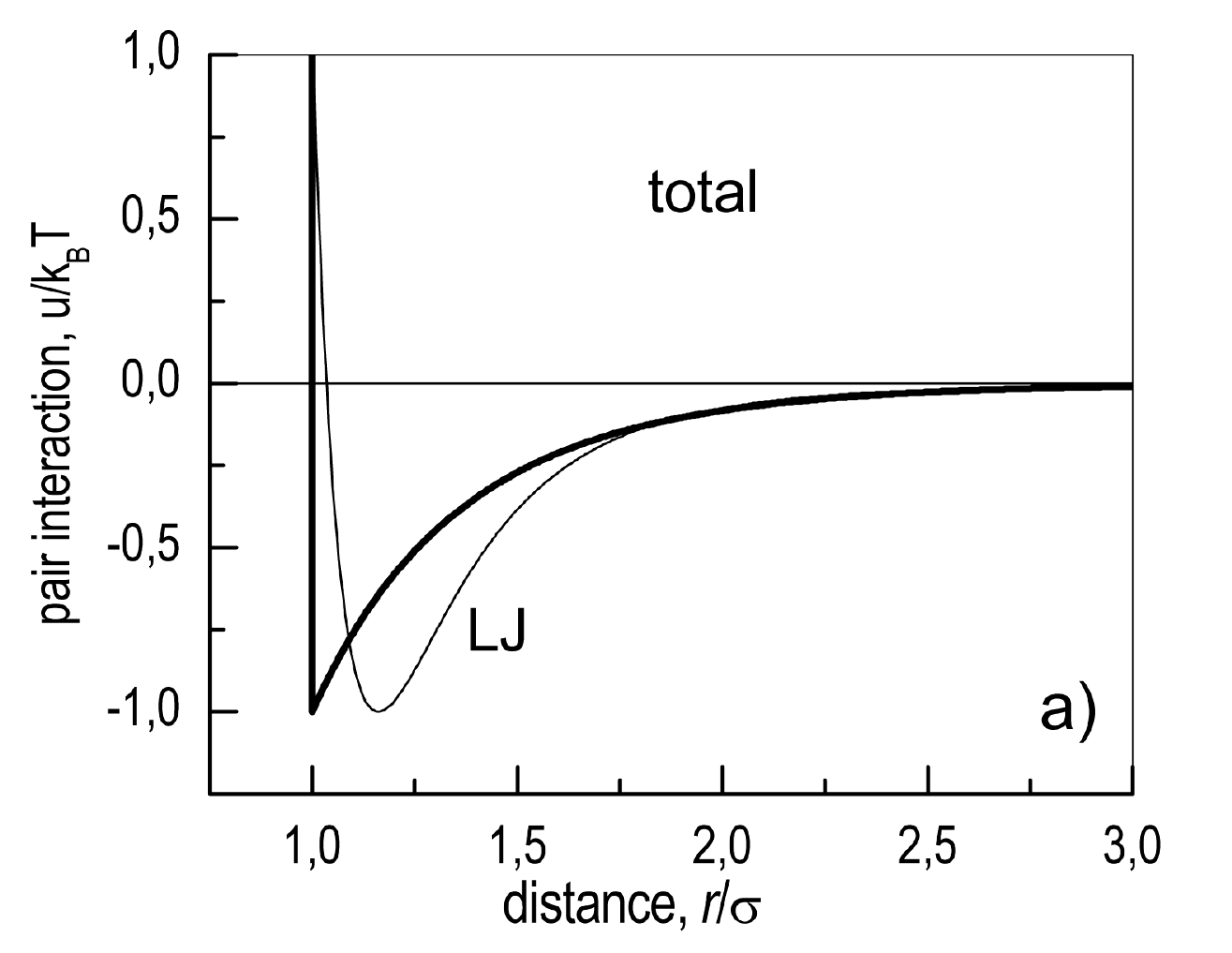}
\includegraphics[width=0.5\textwidth]{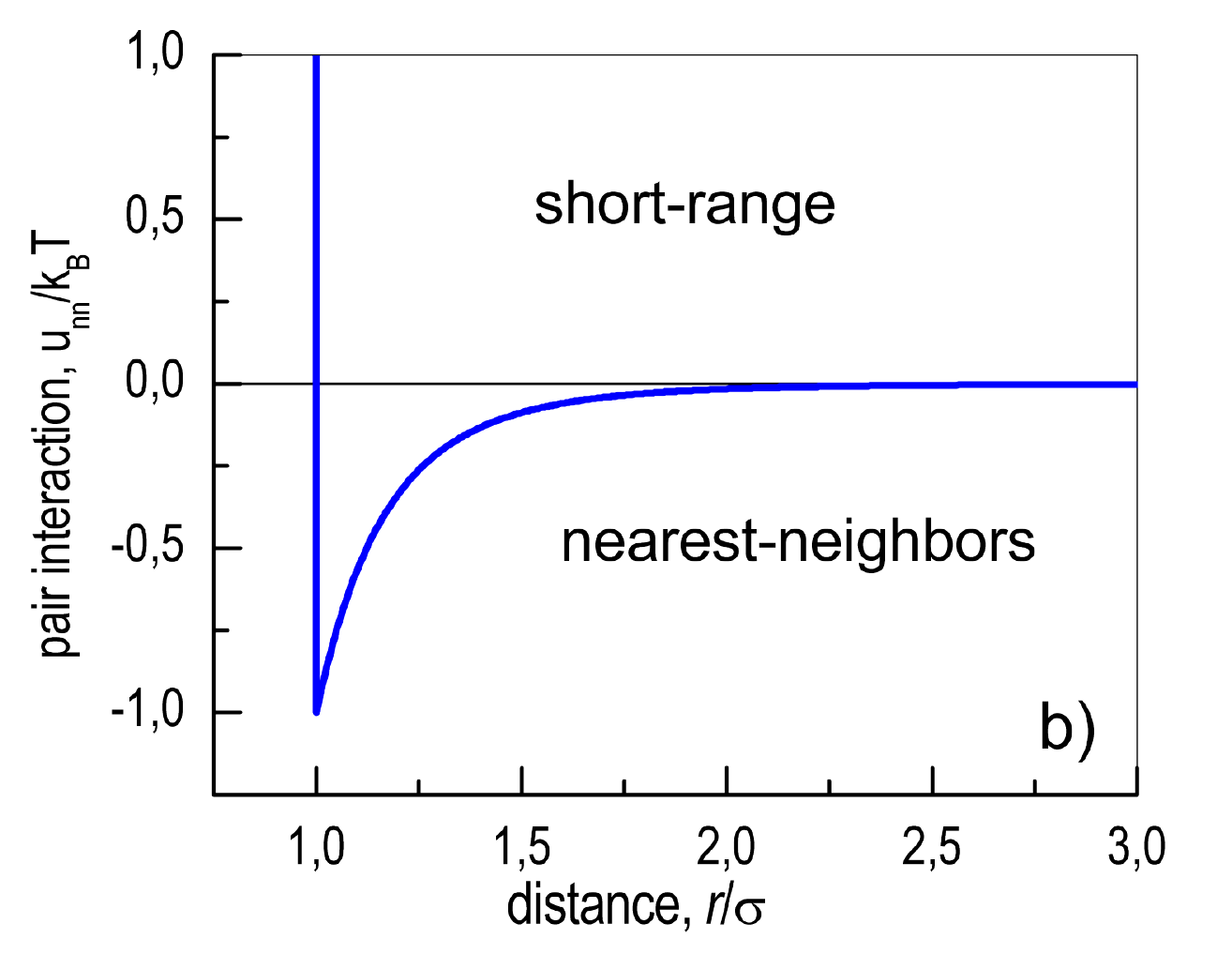}
}
\centerline{
\includegraphics[width=0.5\textwidth]{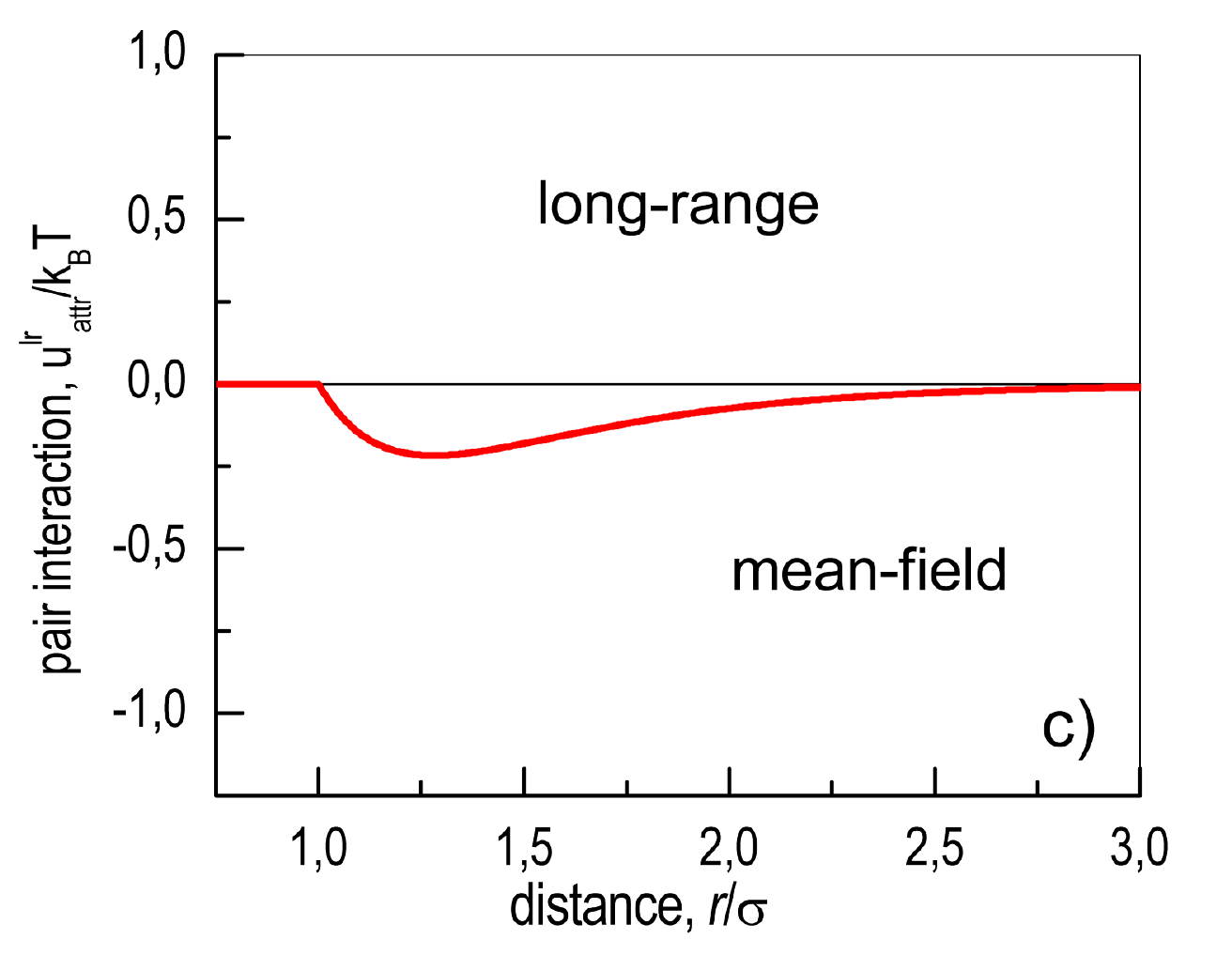}
}
\caption{(Color online) Pair interaction energy  in the LJ-like HCAY fluid, equation~(\ref{uHCAY}), and its comparison against the LJ counterpart (thin solid line)~--- part~(a).
The decomposition of the LJ-like HCAY pair interaction energy into the short-range interaction energy between two neighboring  molecules or excluded volume interaction energy,
equation~(\ref{unn}),  with $z_{0} \sigma=4$  [part~(b)] and the weak long-range attractive interaction energy between target molecule and any other molecule outside the first coordination shell, equation~(\ref{ulr}) [part~(c)]. }
\label{FigPOT}
\end{figure}

In this communication we wish to focus on another popular model system in the liquid state theory, namely, the LJ-like hard-core attractive Yukawa (HCAY) fluid  model \cite{HendersonMP1978}
\begin{eqnarray}
u(r) =
\left\{
\begin{array}{ll}
\infty, & \quad r<\sigma,  \\ \\
-{\displaystyle \frac{ \epsilon\sigma}{ r}}{\re}^{-z(r-\sigma)}, & \quad r \geqslant \sigma
\end{array}
\right.
\label{uHCAY}
\end{eqnarray}
with $z\sigma=1.8$ [see the part~(a) in figure~\ref{FigPOT} for details regarding the relation between the LJ-like HCAY interaction potential and the original LJ potential].
This model fluid has been studied intensively by computer experiment  \cite{Shukla2000,Minerva2001,lomba} as well as by other approaches, such as the mean-spherical integral equation theory (MSA) \cite{HendersonMP1978}, the MSA-based first-order perturbation theory (FMSA) \cite{tang2000}, the MSA-based high temperature expansions theory \cite{blum,duh}.

As for the virial expansion approach,
to the best of our knowledge, there is only one paper by Naresh and Singh~\cite{NareshSingh2009}, where the virial coefficients up to the sixth order, i.e., $B_2(T)$, $B_3(T)$, $\ldots $, $B_6(T)$, for the LJ-like HCAY fluid have been reported.
After substituting these coefficients into the virial EOS, equation~(\ref{Pvir}), the applicability of the latter in the case of the LJ-like HCAY fluid (see figure~\ref{FigZ18} for illustration) can be summarized as follows~\cite{NareshSingh2009}: (i) being truncated by $B_6(T)$, the  virial EOS is rather accurate in the density range up to a reduced density
$\rho\sigma^3 \approx 0.5$, and remains to be qualitatively correct for the entire density range, but only for the reduced temperatures $T^* = k_{\rm B}T/\epsilon=1.5$ and 2 that are supercritical temperatures for this fluid model
(the critical point temperature in this case $T^*_{\rm c} \approx 1.2$ \cite{Shukla2000,Minerva2001,MelnykFPE2009});
(ii) the same virial EOS begins to fail already right after the density $\rho\sigma^3 \approx 0.1$ in the case of the subcritical temperatures (the results shown by the dashed line in figure~\ref{FigZ18} for the reduced temperature $T^* =1$).

By using the LJ-like HCAY model fluid as a pilot system, the purpose of this exploratory study is to show that even in the case of the truncated virial EOS, the performance  of the latter in the wide range of density and temperature conditions,
including the subcritical ones, can be substantially improved by implementing the ideas that were elaborated within the framework of the augmented van der Waals theory \cite{MelnykFPE2009,MelnykJSF2010}.
These ideas concern the issue of a split of the total interaction potential $u(r)$ into two terms.
Namely,  in contrast to presumably  van der Waals's suggestion that the total potential energy is composed of the
repulsion and attraction contributions, the ``augmented'' version of the van der Waals theory means that one term is representing the  most accurate possible the full excluded volume energy in the system, that is the interaction energy between the neighbouring molecules,
while the remaining part is responsible for the weak long-range attractive interaction energy,  or the energy of cohesion, between the next-neighbouring molecules.

The remainder of this paper is organized as follows: in section~\ref{sec:2}, we  provide an overview of the augmented van der Waals theory \cite{MelnykFPE2009,MelnykJSF2010}. In section~\ref{sec:3}, we discuss how the ideas of this theory can be implemented within the virial expansion approach and we present the corresponding results for the LJ-like HCAY fluid in section~\ref{sec:4}. We conclude with section~\ref{sec:5}.

\section{Augmented van der Waals theory}
\label{sec:2}

Recently, it has been shown that the thermodynamics as well as the vapour--liquid equilibrium in the LJ-like HCAY fluid can be rather accurately described within the augmented van der Waals theory~\cite{MelnykFPE2009,MelnykJSF2010}.
In particular, within this theory, the EOS of the LJ-like HCAY model fluid reads,
\begin{equation}
p  = p_{\rm 0}  - \rho^2 a ,
\label{pvdw}
\end{equation}
where, similar to the original van der Waals suggestion,  coefficient $a$ is related to the contribution from the attractive interaction energy between molecules, while pressure $p_0$ stands for the  pressure due to the excluded volume energy.
Following van der Waals, the excluded volume pressure $p_{\rm 0}$
originates from the fact that in the system of a volume $V$ and composed of $N$ molecules with a hard-core diameter $\sigma$, each molecule excludes an amount of volume $v_{\rm 0}$ from being allowed to explore by all other molecules of the system.
Thus, the volume accessible for molecules  is reduced to $V-Nv_{\rm 0}$. This phenomenon, that firstly was pointed out by van der Waals, results in the pressure,
\begin{equation}
p_{\rm 0}
= \frac{N k_{\rm B}T}{V - N v_{\rm 0}} = \frac{\rho k_{\rm B}T}{1 - \rho v_{\rm 0}}\,,
\label{pEXCL}
\end{equation}
that is referred to as the excluded volume pressure.

\begin{figure}[!h]
\includegraphics[width=0.31\textwidth]{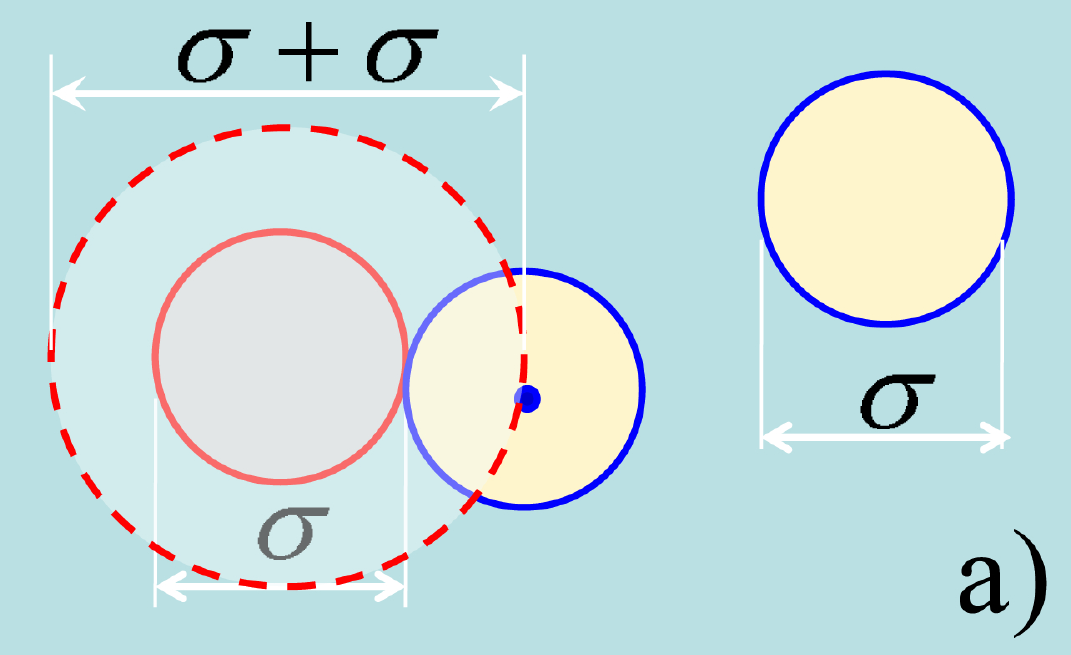}
\hspace{0.2cm}
\includegraphics[width=0.31\textwidth]{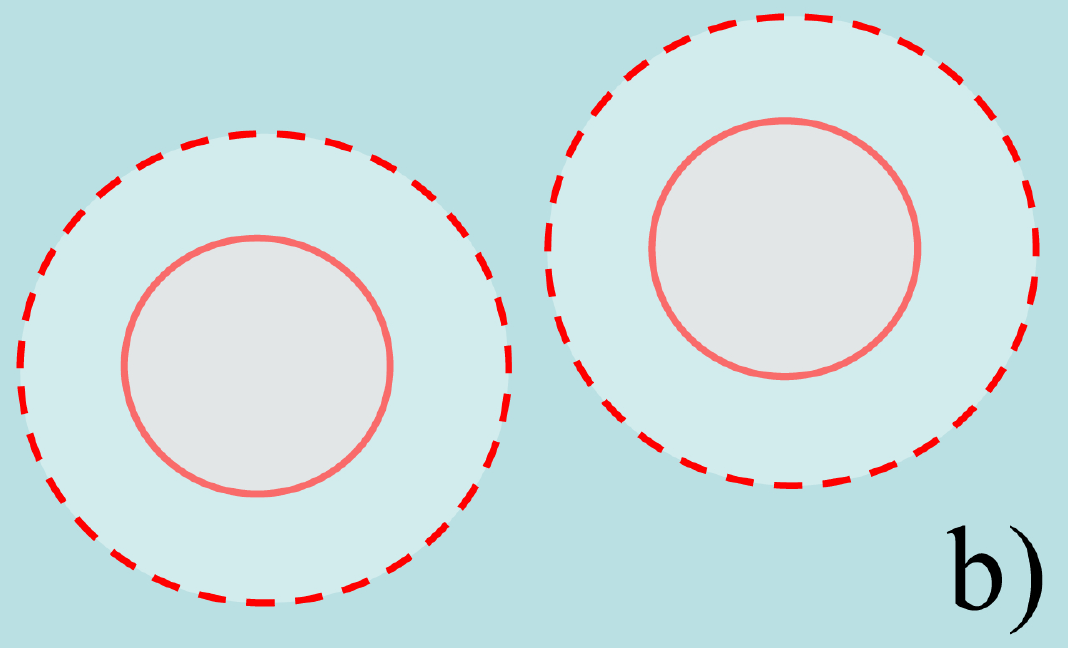}
\hspace{0.2cm}
\includegraphics[width=0.31\textwidth]{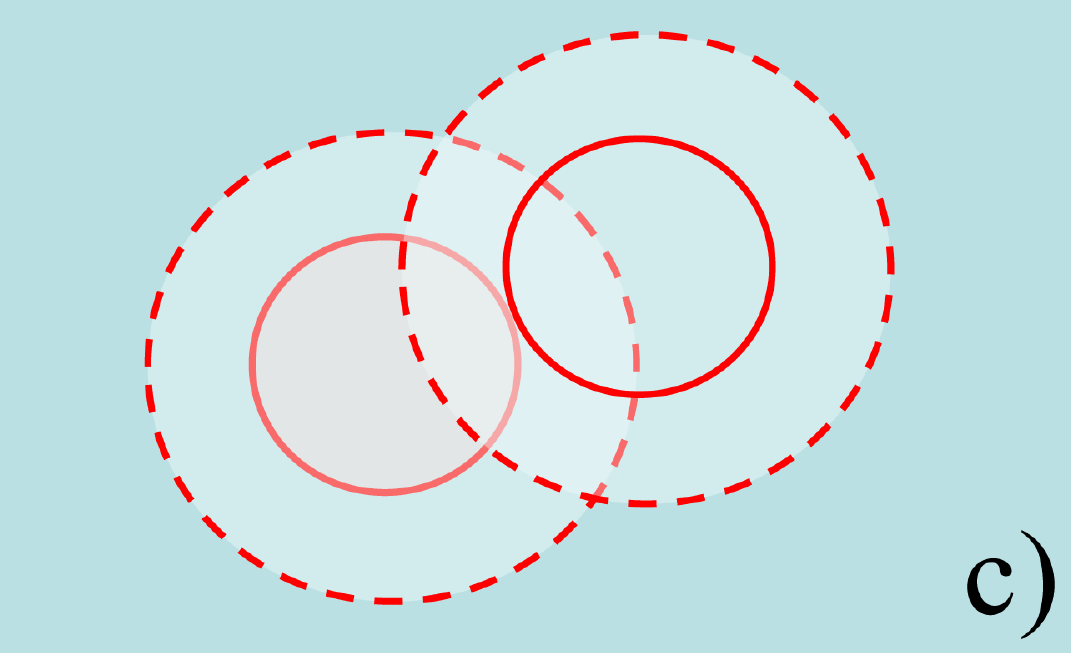}
\caption{
Towards definition of the excluded volume in a fluid composed of molecules with a hard core diameter $\sigma$, and its dependence on the distance between the pair of the nearest-neighbour molecules.
}  \label{FigEXCL}
\end{figure}

The excluded volume itself is
uniquely defined upon the distance between each pair of the neighbouring molecules in the fluid  (see figure~\ref{FigEXCL}).
As the first approximation, van der Waals assumed that the excluded volume per molecule is a constant and equals fourfold the molecular volume, i.e., $v_{\rm 0} = b \equiv (2/3)\pi\sigma^3$.
Indeed, it is the case for a dilute gaseous phase   [see figure~\ref{FigEXCL}~(b)] when the mean distance $\langle r \rangle$ between molecules is large (more precisely, when the mean distance $\langle r \rangle$ between the centers of the pair of the neighbouring molecules is larger than $2\sigma$).
In the dense gaseous phase and, especially, in the liquid phase, the mean distance $\langle r \rangle$ between the neighbouring  molecules becomes
shorter than $2\sigma$, and, consequently, the excluded volume shells start to overlap [see figure~\ref{FigEXCL}~(c)], resulting in the excluded volume per molecule being smaller than fourfold the molecular volume, i.e., $v_{\rm 0} < b$.

%

Since it is rather evident that
the mean intermolecular distance $\langle r \rangle$ between the neighbouring mole\-cules
is affected by the number density $\rho$, it became necessary to incorporate the density dependence into the excluded volume $v_{\rm 0}$ as well.
The most natural way to comply with this
requirement was to utilize the hard-sphere fluid model  for the evaluation of the excluded volume contribution to the whole spectrum  of fluid properties and to the EOS, in particular.
Such an assumption lies behind
the perturbation theory of fluids due to Zwanzig \cite{Zwanzig} and was successfully exploited by Widom \cite{WidomSCI1963}, Barker and Henderson \cite{BH1976},
Weeks, Chandler and Andersen \cite{wcaSCI1980} and many others \cite{HM2013}.

However, although it is less evident, but the mean intermolecular distance $\langle r \rangle$ between two neighbouring molecules is affected by the temperature as well.
Such a feature of the intermolecular distance is mediated by the energy of the short-range attractive interaction energy
that together with the energy of the hard-sphere repulsion
are present for the pair of neighbouring molecules.
This observation suggests that the short-range repulsive and short-range attractive interaction energies between the neighbouring molecules should
be incorporated into the scheme to evaluate the contribution of the excluded volume to the pressure. 

As the first step to comply with this idea, let us follow the suggestion \cite{MelnykFPE2009,MelnykJSF2010} and present the pair interaction energy $u(r)$
in the form,
\begin{equation}
u(r) =
u_{\rm nn}(r) + u^{\rm lr}_{\rm attr}(r)\,,
\label{utmn}
\end{equation}
which is in contrast to the common practice~\cite{Zwanzig,WidomSCI1963,BH1976,wcaSCI1980}
that prefers to utilize another form,
\begin{equation}
u(r) =
u^{\rm hs}(r) + u^{\rm }_{\rm attr}(r)\,,
\label{uths}
\end{equation}
which assumes that the pair interaction energy $u(r)$ is separated into purely repulsive hard-sphere term $u^{\rm hs}(r)$ and attractive interaction energy $u_{\rm attr}(r)$ contributions.
The interaction energy $u_{\rm nn}(r)$ in equation~(\ref{utmn}) represents the full interaction energy of a target molecule and its neighbouring (nn) counterpart.
The neighbouring molecules
and the corresponding interaction energy $u_{\rm nn}(r)$ are identified by means of the range (distance) criterion. According to this criterion, the excluded volume
interaction energy $u_{\rm nn}(r)$
includes the full energy of the hard-core repulsion $u^{\rm hs}(r)$ and  only a part
of the full attraction energy,
namely, the part that is responsible for the interaction
of a target molecule with its nearest-neighbour molecule only,
\begin{eqnarray}
\label{unn}
u_{\rm nn}(r) &\equiv& u^{\rm hs}(r) + u^{\rm sr}_{\rm attr}(r)
=\left\{
\begin{array}{ll}
\infty, &  r<\sigma,  \\ \\
-{\displaystyle \frac{\epsilon\sigma}{r}} \re^{-z_{0}(r-\sigma)}, &  r\geqslant \sigma.
\end{array}
\right.
\end{eqnarray}
In fact, this is the interaction energy with the molecules that belong to the first coordination shell of a target molecule. Following such a definition,
the attraction $u^{\rm sr}_{\rm attr}(r)$ incorporates
the full attraction energy $u^{\rm}_{\rm attr}(r)$ at the contact distance $r=\sigma$ between two molecules, but decays  faster than the full attraction energy, in order not to exceed the radii of the first coordination shell.
In reality, the range of the short-range attraction $u^{\rm sr}_{\rm attr}(r)$ is around one molecular hard-core diameter $\sigma$, and in the present case can be approximated by fixing the decay parameter at $z_{0}\sigma = 4$.
Then, the term $u^{\rm lr}_{\rm attr}(r)$ in equation~(\ref{utmn})  is determined as the difference, $u(r) - u_{\rm nn}(r)$, and reads
\begin{eqnarray}
\label{ulr}
  u^{\rm lr}_{\rm attr}(r) \equiv u(r) - u_{\rm nn}(r)
= \left\{
 \begin{array}{llll}
0, &  r\leqslant\sigma,  \\ \\
-{\displaystyle\frac{\epsilon\sigma}{r}}\left[\re^{\displaystyle -z(r-\sigma)} - \re^{\displaystyle -z_{0}(r-\sigma)}\right], &  r > \sigma.
\end{array}
\right.
\end{eqnarray}
The pair potential $u^{\rm lr}_{\rm attr}(r)$ corresponds to the interaction energy of the target molecule with any other molecule but from outside the first coordination shell.
Figure~\ref{FigPOT} shows the total pair interaction energy $u(r)$, the excluded volume interaction energy $u_{\rm nn}(r)$,
and the long-range attractive interaction energy $u^{\rm lr}_{\rm attr}(r)$, all according to their definitions by equations~(\ref{uHCAY}), (\ref{unn}) and (\ref{ulr}), respectively.

The shape  of the long-range attractive interaction energy $u^{\rm lr}_{\rm attr}(r)$ [see figure~\ref{FigPOT}~(c)]
appears to be crucial~\cite{MelnykJSF2010} for the evaluation of the van der Waals coefficient $a$,
that in general case is given by
\begin{equation}
a = -2\pi\int_0^\infty g^{0}(r)u^{\rm lr}_{\rm attr}(r)r^2\rd r.
\label{aexact}
\end{equation}
The function $g^{0}(r)$ in this equation  stands now for the radial distribution function of the system with the excluded volume interaction potential $u_{\rm nn}(r)$.
We wish to stress, that only by using  for excluded volume interaction energy $\,u_{\rm nn}(r)$ its definition according to equation~(\ref{unn}),
it is possible to justify the so-called mean-field assumption, $g^{0}(r)=1$,
in equation~(\ref{aexact}). In the case of the LJ-like HCAY fluid, this results in the simple expression,
\begin{eqnarray}
\label{avdw}
a = -2\pi\int_0^\infty u^{\rm lr}_{\rm attr}(r)r^2\rd r
= -2\pi\sigma^3\epsilon \left(\frac{1+z\sigma}{z^2\sigma^2} -
\frac{1+z_{0}\sigma}{z_{0}^2\sigma^2}\right)\,.
\end{eqnarray}

\begin{figure}[!t]
\centerline{
\includegraphics[width=0.47\textwidth]{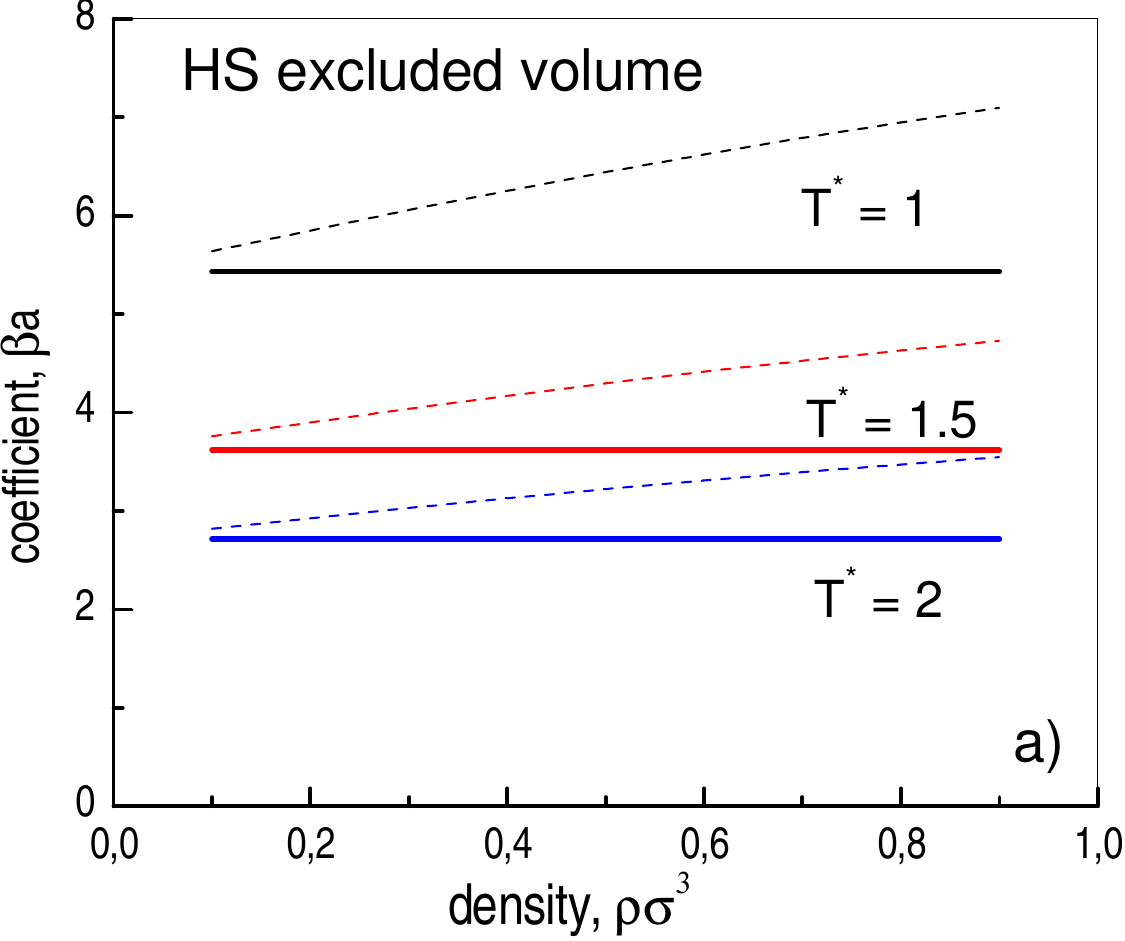}
\hspace{0.5cm}
\includegraphics[width=0.47\textwidth]{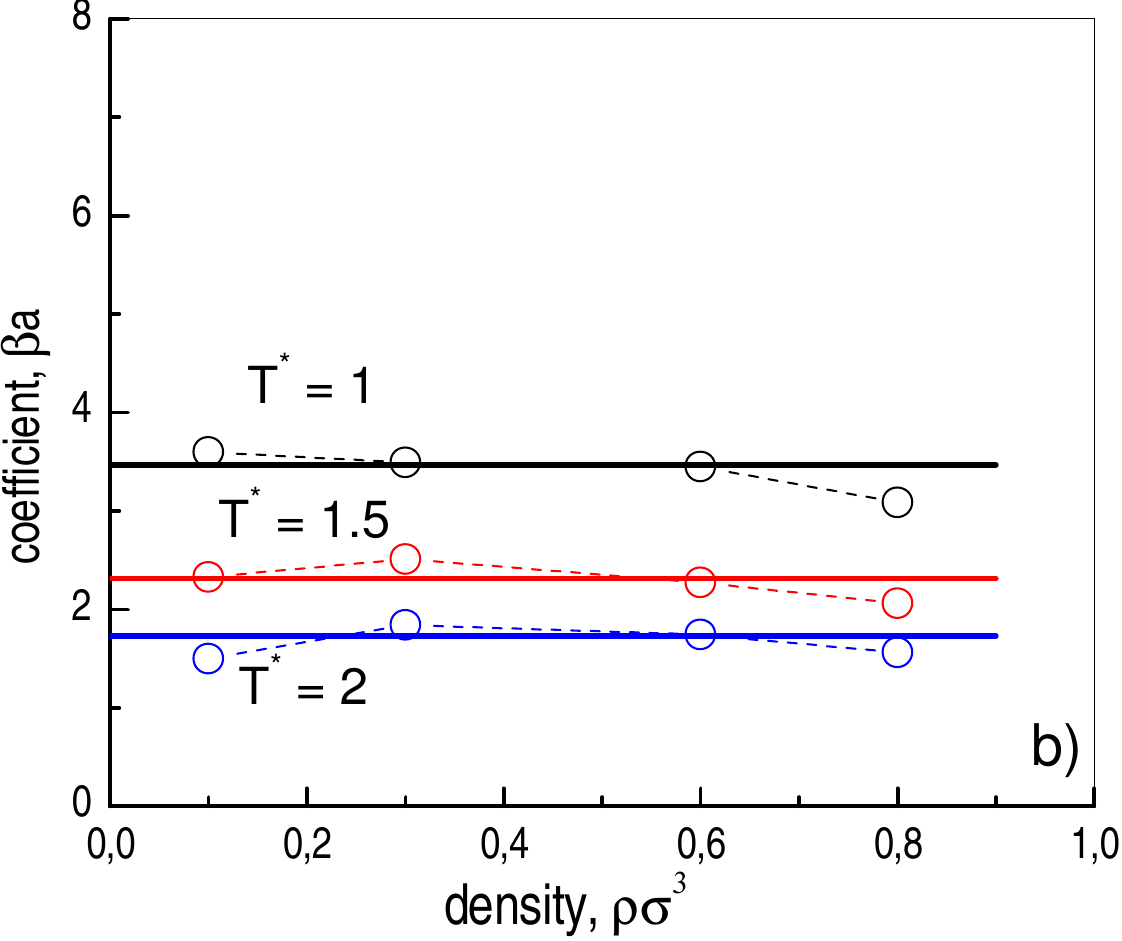}
}
\caption{(Color online)
Coefficient $a$ of the augmented van der Waals EOS (\ref{pvdw}) as calculated in accordance with its definition by equation~(\ref{aexact}) (dashed line) and within the mean-field approximation, $g^{0}(r)=1$, given by equation~(\ref{avdw}) (solid line) for two different choices of the excluded volume model for the same LJ-like HCAY fluid: (i) excluded volume is described within the hard-sphere model~--- part~(a), and (ii) within the short-range Yukawa model~--- part~(b).
}
\label{FigCoefa}
\end{figure}

Figure~\ref{FigCoefa} shows the results for coefficient $a$ as they are obtained for two different choices of the excluded volume model and, consequently, for two different energies of the long-range attractive interaction energy, $u^{\rm}_{\rm attr}(r)$ and $u^{\rm lr}_{\rm attr}(r)$, in the case of the same LJ-like HCAY fluid. Namely, figure~\ref{FigCoefa}~(a) corresponds to the case when the excluded volume is described within the traditional hard-sphere model [in accordance with equation~(\ref{uths})], while figure~\ref{FigCoefa}~(b) corresponds to the case when the excluded volume is described within the proposed short-range attractive Yukawa model [in accordance with equations~(\ref{utmn}) and (\ref{unn})].
Two values for the coefficient $a$ in the case of each of these models were evaluated: (i) by numerical integration, in accordance with the definition by equation~(\ref{aexact}), and (ii) analytically, within the mean-field approximation,  equation~(\ref{avdw}).
We note, when calculating the integral in
equation~(\ref{aexact}), that we have used for the radial distribution function $g^{0}(r)$ the closed-form analytical equation~\cite{adtjcp2005,adtjcp2005_2} in the case of hard-sphere model, and the Monte Carlo simulation data~\cite{MelnykFPE2009} in the case of short-range attractive Yukawa model.

Obviously, the magnitude of the coefficient $a$ is different for each model. However, the most intriguing insight from figure~\ref{FigCoefa} comes from analysing the values of coefficient $a$
within the same model but obtained from two different equations, equations~(\ref{aexact}) and (\ref{avdw}), respectively. We can see that in the case of the hard-sphere model for excluded volume interaction [figure~\ref{FigCoefa}~(a)], the mean-field and exact values of coefficient $a$ are quite different, both quantitatively and qualitatively. By contrast, in the case of
short-range attractive Yukawa model, we could admit the tendency for coefficient $a$ to be the same, independently of equations, (\ref{aexact}) or (\ref{avdw}), and, consequently, which is most important~---  to be independent of the density.

\section{Augmented virial EOS}
\label{sec:3}

Augmented virial EOS, which is one of the goals of the present study, will be obtained as the series
in powers of the density $\rho$ of the augmented EOS (\ref{pvdw}).
Since  coefficient $a$ does not depend on density within the augmented van der Waals theory,  the series will concern exclusively the excluded volume pressure $p_{\rm 0}$.
Namely,
%
\begin{eqnarray}
\label{PexclVir}
\frac{p_{\rm 0}}{k_{\rm B}T} \equiv \frac{p^{\rm nn}}{k_{\rm B}T} =
 \rho + B^{\rm nn}_2(T)\rho^2 + B^{\rm nn}_3(T)\rho^3 + B^{\rm nn}_4(T)\rho^4 + \ldots \, ,
\end{eqnarray}
where the virial coefficients $B^{\rm nn}_n(T)$ are defined exactly as it is discussed in equations (\ref{B2})--(\ref{fM}), but instead of the total interaction energy $u(r)$, the short-range excluded volume interaction energy $u_{\rm nn}(r)$
must be used in the exponential of the Mayer function.

The resulting augmented virial EOS
is obtained by substituting the virial series for the excluded volume pressure, equation~(\ref{PexclVir}),
into the augmented van der Waals EOS, equation~(\ref{pvdw}),
\begin{eqnarray}
\label{PaugVir}
\frac{p}{k_{\rm B}T} &=& \rho + \left(\frac{}{}B^{\rm nn}_2(T) + \frac{a}{k_{\rm B}T}\right)\rho^2  
+ B^{\rm nn}_3(T)\rho^3 + B^{\rm nn}_4(T)\rho^4 + \ldots\,.
\end{eqnarray}
From the first glance at equation~(\ref{PaugVir}), one  immediately notices two
important features concerning the role that the long-range attraction energy $u^{\rm lr}_{\rm attr}(r)$ plays within the augmented virial expansion approach.
First of all, since
coefficient $a$ does not depend on the density,
it follows that the long-range attraction energy $u^{\rm lr}_{\rm attr}(r)$ contributes  to the second virial coefficient only,
\begin{equation}
B_2(T) = B^{\rm nn}_2(T) + \frac{a}{k_{\rm B}T}\,.
\label{B2aug}
\end{equation}
Secondly,  the remaining augmented virial coefficients $B_3(T)$, $B_4(T)$, $\ldots$ all do not contain the
contribution from the long-range attractive interaction energy $u^{\rm lr}_{\rm attr}(r)$, being identical to those that correspond to the excluded volume or short-range interaction energy $u_{\rm nn}(r)$ only, i.e.,
\begin{equation}
B_n(T) = B^{\rm nn}_n(T)\, \qquad \mbox{for} \qquad n>2\,.
\label{Bnaug}
\end{equation}
In what follows, we apply the augmented virial EOS, equation~(\ref{PaugVir}),
to calculate the compressibility factor $pV/(Nk_{\rm B}T)$ of  the LJ-like HCAY fluid.

\begin{figure}[!b]
\centerline{
\includegraphics[width=0.55\textwidth]{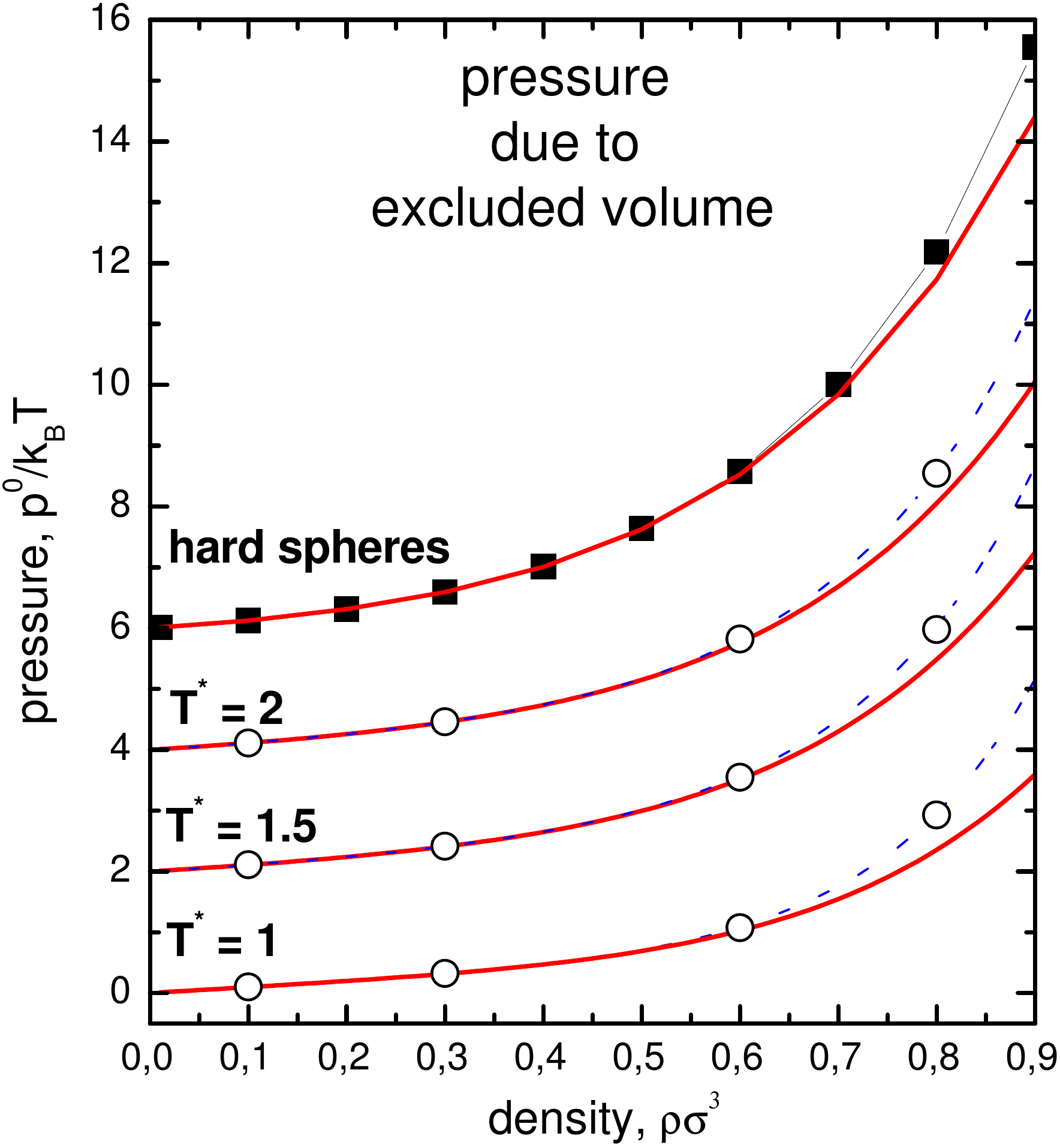}
}
\caption{(Color online)
 Excluded volume pressure, $p_0/(k_{\rm B}T)\equiv p^{\rm nn}/(k_{\rm B}T)$,
 as it is modelled by the hard-core short-range Yukawa attraction energy~(\ref{unn})
 with decay parameter $z_{0}\sigma = 4$.
 The thick solid lines represent results of the  virial EOS (\ref{PexclVir})  truncated at the sixth virial coefficient with coefficients $B_2^{\rm nn}(T)$, $\ldots $, $B_6^{\rm nn}(T)$ reported by Naresh and Singh \cite{NareshSingh2009}, while open circles are the computer experiment data by Shukla~\cite{Shukla2000}. The thin dashed lines connect the symbols and are shown to guide the eye.
The filled squares (computer experiment data) at the top and thin solid line (results  of the virial EOS  truncated at the tenth virial coefficient~\cite{ClisbyMcCoy2006}),
both correspond for the pressure, $p_0/(k_{\rm B}T)\equiv p^{\rm hs}/(k_{\rm B}T)$, of the hard-sphere fluid that represents here the high-temperature ($T^*\to\infty$) limit of the the excluded volume pressure, and is shown here for comparison purposes; the thick solid this case, like in all other cases in this figure, represents the virial EOS of the hard-sphere fluid being truncated at the sixth virial coefficient. The pressure isotherms have been shifted for clarity.
}
 \label{Fig3}
\end{figure}

\section{Results and discussions}
\label{sec:4}

Up to date, only the first five virial coefficients, $B^{\rm nn}_{2}(T)$, $\ldots$, $B^{\rm nn}_{6}(T)$,
for the excluded volume interaction energy $u_{\rm nn}(r)$  are known~\cite{NareshSingh2009}.
Figure~\ref{Fig3} shows a set of data for the excluded volume
pressure, $p_{\rm 0} \equiv p^{\rm nn}$, that result from the virial EOS,  equation~(\ref{PexclVir}), (solid lines) truncated
at the sixth virial coefficient
as well as those that were obtained from computer experiment (symbols) to compare.
There are three isotherms, namely, $T^*=1, 1.5$ and 2 that correspond to the excluded volume pressure within the short-range attraction Yukawa model, while
the forth isotherm represents the pressure of the hard-sphere fluid, i.e. corresponds to the case  when $T^*\to\infty$.
The most important conclusions that follow from the results presented in figure~\ref{Fig3} concern the accuracy and, perhaps, even more generally~--- applicability of the virial expansions approach in the case of excluded volume interactions. First of all, we can see that  virial EOS,  equation~(\ref{PexclVir}), being truncated at the sixth virial coefficient, reproduces rather accurately the data from computer experiment in the density range  $0 < \rho\sigma^3 < 0.6$. We note, that this observation practically does not depend on the temperature; similar behaviour is found in the case  of the hard-sphere model as well, if the virial series is truncated at the sixth virial coefficient. At the same time, by analysing the results of the hard-sphere model we can
suggest, that truncation of the virial EOS, equation~(\ref{PexclVir}), at the tenth virial coefficient must
be sufficient to provide rather accurate description of the excluded volume
pressure, $p_{\rm 0} \equiv p^{\rm nn}$, of the LJ-like HCAY fluid in the full density range.
\begin{figure}[!t]
{\includegraphics[width=0.45\textwidth]{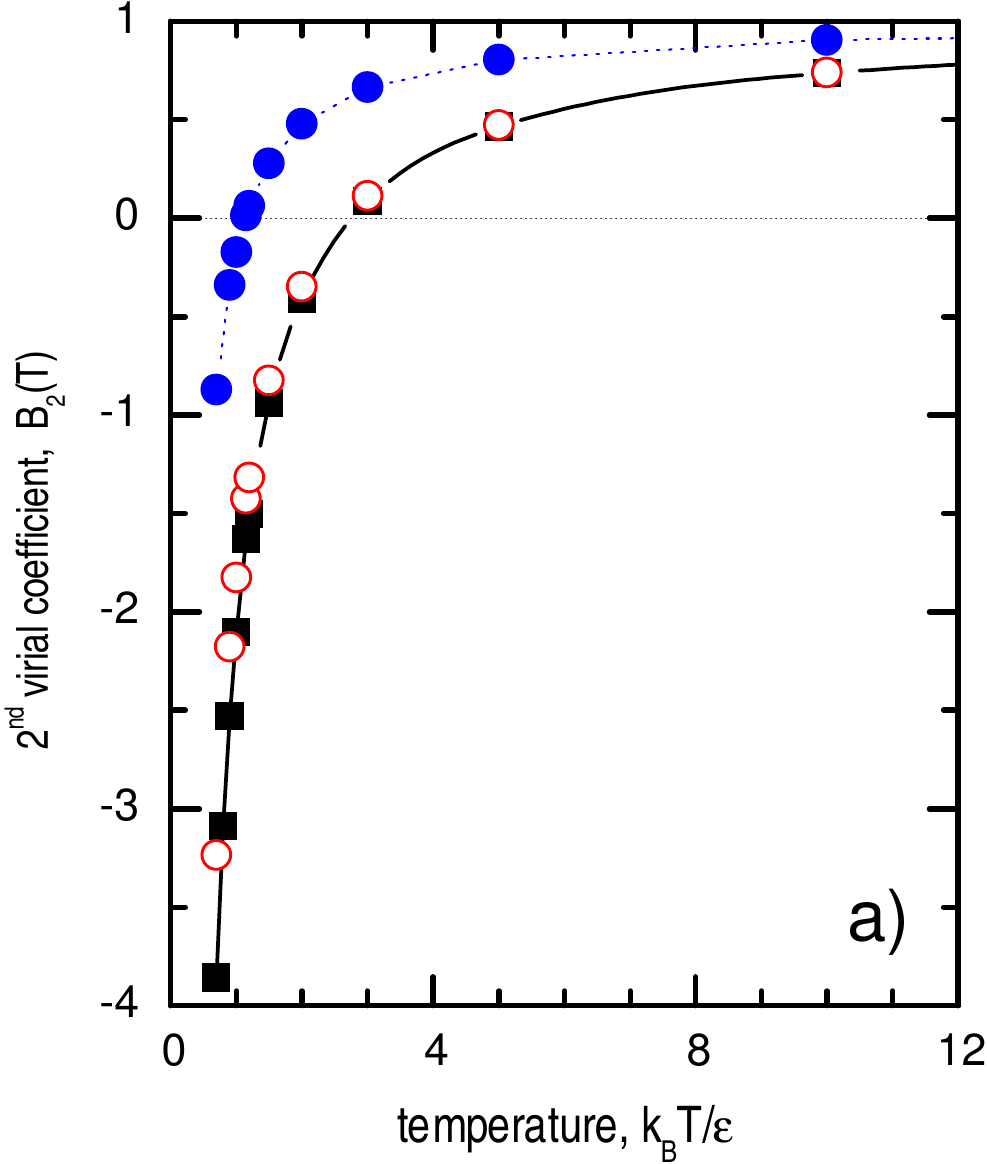}}
\hspace{0.5cm}
{\includegraphics[width=0.45\textwidth]{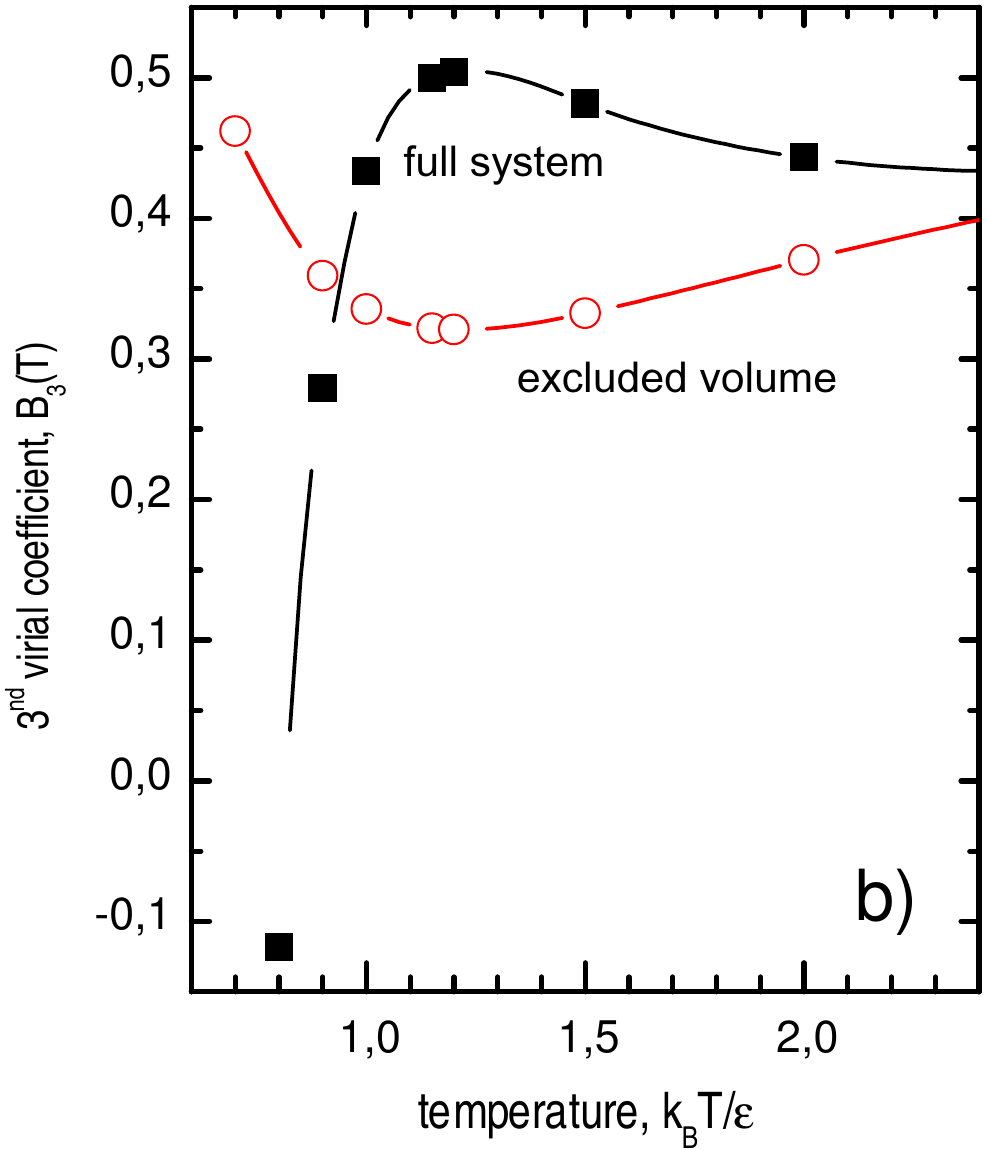}}
\caption{(Color online) The second and third virial coefficients of the LJ-like HCAY fluid.  The filled squares with a connecting line represent the results for $B_2(T)$ and $B_3(T)$ that were obtained by Naresh and Singh \cite{NareshSingh2009} using equations~(\ref{B2}) and (\ref{B3}), respectively.
The open circles with a connecting line represent the results for $B_2(T)$ and $B_3(T)$ that correspond to the augmented virial expansion approach and are defined in accordance with equations~(\ref{B2aug}) and (\ref{Bnaug}), respectively.
The filled circles in part a show results for the second virial coefficient that are obtained in accordance with equation~(\ref{B2aug}) but for the case when the pair interaction energy, $u(r)$, is separated into two parts following a common practice \cite{BH1976,Zwanzig,WidomSCI1963,wcaSCI1980}, i.e., it consists of purely repulsive hard-sphere energy, $u^{\rm hs}(r)$, and full attractive interaction energy, $u_{\rm attr}(r)$.
 }
 \label{FigB3B2}
\end{figure}

The definition of the augmented virial coefficients $B_n(T)$ in accordance to equations~(\ref{B2aug}) and (\ref{Bnaug}) differs from a rigorous that [e.g., see equations~(\ref{B2}) and (\ref{B3})].
Obviously, one would expect the different values for virial coefficients $B_n(T)$ from these two definitions. Indeed, this is the case for the virial coefficients $B_3(T)$, $\ldots$, $B_6(T)$, i.e., for $B_n(T)$ with $n>2$. As an example, figure~\ref{FigB3B2}~(b) shows the results for $B_3(T)$, where the augmented virial coefficient  differs significantly from its conventional counterpart, remaining positive even for extremely low temperature.  However,  it does not in the case of the second virial coefficient when two definitions, given by equations~(\ref{B2}) and (\ref{B2aug}), both result in practically the same values of $B_2(T)$ as it is illustrated in figure~\ref{FigB3B2}~(a). In particular, the Boyle temperature [the temperature at which $B_2(T)$ assumes zero value] of the LJ-like HCAY fluid  in both cases is the same, being fixed at approximately $T^*_\textrm{B}\sim 2.7$.
We note, that this feature of the second virial coefficient is sensitive to the way how the total pair interaction $u(r)$ is split into the excluded volume and long-range contributions. To illustrate this point, figure~\ref{FigB3B2}~(a) shows the results for the augmented $B_2(T)$ in the case when the $u(r)$ is split in accordance to the common practice \cite{Zwanzig,WidomSCI1963,BH1976,wcaSCI1980} given by equations~(\ref{uths}), i.e., when
the nearest-neighbour interaction potential $u_{\rm nn}(r)$ consists of the hard-core repulsion $u^{\rm hs}(r)$ only. We can see, that second virial coefficient in this case is quite different from a rigorous one.

\begin{figure}[!t]
\centerline{
\includegraphics[width=0.55\textwidth]{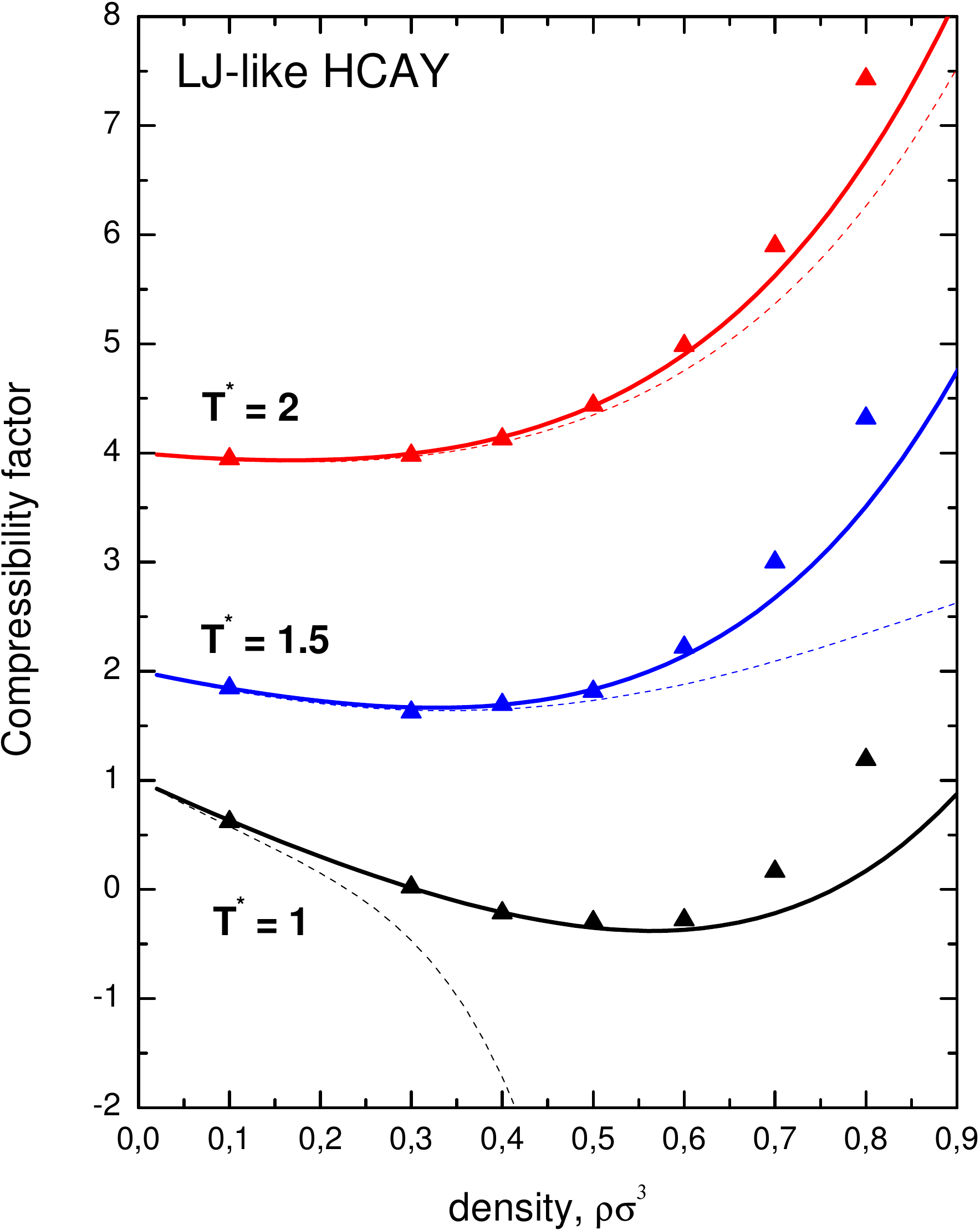}
}
\caption{(Color online) Compressibility factor $pV/(Nk_{\rm B}T)$
of the LJ-like HCAY fluid. The solid lines represent the results of the truncated augmented virial EOS, equation~(\ref{PaugVir}), truncated at the sixth virial coefficient, while symbols correspond to the computer experiment data by Shukla \cite{Shukla2000}. The dashed lines show the results of the truncated conventional virial EOS, equation~(\ref{Pvir}),
that are taken from the study by Naresh and Singh~\cite{NareshSingh2009}. The temperature conditions are specified in the figure. We note that critical point temperature for the LJ-like HCAY fluid is estimated to be around $T^*_{\rm c} \approx 1.2$ \cite{Shukla2000,Minerva2001,MelnykFPE2009}.
The curves have been shifted for clarity. }
 \label{FigZ18}
\end{figure}

The results for compressibility factor, $pV/(Nk_{\rm B}T)$,
of the LJ-like HCAY fluid, that follow from the augmented virial EOS (\ref{PaugVir}), are shown  in figure~\ref{FigZ18} (the thick solid lines) to be compared against computer experiment data~\cite{Shukla2000} as well as against the rigorous virial EOS (\ref{Pvir}).
We can see that for all three temperatures that include both the supercritical and subcritical conditions, there are notable improvements in searching for an agreement with computer experiment data.
However, the most valuable result concerns the performance of the augmented virial EOS (\ref{PaugVir}) at the subcritical temperature  $T^*=1$, where the original virial EOS  equations~(\ref{Pvir}) fails.
Some discrepancies between the augmented virial EOS and computer experiment data, that still are observed for densities $\rho\sigma^3 > 0.6$, are pretty similar to those that we already discussed in figure~\ref{Fig3}, and they should be attributed to the truncation of the excluded volume virial EOS (\ref{PexclVir}) at the sixth virial coefficient.

\section{Conclusions}
\label{sec:5}

In the present study, the issue of the performance of the  virial expansion approach in the liquid state theory is discussed by using as an example the LJ-like HCAY fluid model.
More precisely, we were devoted here to discuss the issue of the divergence of the  conventional virial EOS (\ref{Pvir}) at subcritical conditions that
for the LJ-like HCAY fluid recently was reported by Naresh and Singh~\cite{NareshSingh2009}.

To deal with this issue, the recent advances~\cite{MelnykFPE2009,MelnykJSF2010} in the augmented version of the van der Waals theory have been explored.
The essence of the van der Waals theory lies in the non-trivial split of total interaction energy into two parts which are forced to be responsible for the excluded volume energy and the cohesive energy, respectively. Traditionally, in what is called ``van der Waals picture of liquids'' \cite{HM2013,Zwanzig,WidomSCI1963,BH1976,wcaSCI1980} it is assumed that excluded volume part is well represented by the hard-core repulsion energy, while the cohesive part is associated with the remaining full attractive interaction energy.
By contrast, within the augmented van der Waals theory~\cite{MelnykFPE2009,MelnykJSF2010}, it considers that not only the repulsive part, but the full energy of interaction between the pair of neighbouring molecules, must be treated as the excluded volume energy, and what remains is representing the cohesion energy.

By applying the virial expansion approach to the augmented van der Waals EOS (\ref{pvdw}), it is obtained that the cohesion energy, which is the long-range part
of the total interaction energy,
contributes to the second virial coefficient, $B_2(T)$, only. All other virial coefficients result from the short-range excluded volume interaction energy, $u_{\rm nn}(r)$, which, however,  consists of both the hard-core repulsion energy and short-range attraction energy between the neighbouring molecules.

To define the excluded volume interaction energy, $u_{\rm nn}(r)$, we have used in this study
the distance
criterion, requiring that the range of excluded volume interaction should not exceed the molecular hard-core diameter $\sigma$.
However, from the analysis of the second virial coefficient in figure~\ref{FigB3B2}~(a) we may conclude that as the criterion for how much of attraction energy in the total pair interaction energy, $u(r)$, should be included into the excluded volume term, $u_{\rm nn}(r)$,
might be the requirement of
the equality between
the augmented second virial coefficient, equations~(\ref{B2aug}), and the rigorous one, equation~({\ref{B2}}).  Namely, the Boyle temperature that follows from the augmented van der Waals theory should be pretty close to the true one, which follows from the rigorous second virial coefficient; otherwise the physics of two systems could differ as well.

The range of attractive interaction energy
seems to be an important issue for the convergence of the virial expansion approach.
For example, in the limiting case of the hard-sphere repulsion, $u^{\rm hs}(r)$, when there is no attractive tail at all, the virial EOS shows no sign of divergence
(e.g., see reference~\cite{Masters2008} and discussion therein).
Very similar conclusions can be drawn  for the  virial EOS (\ref{PexclVir}) truncated at the sixth virial coefficient in the case of model fluid defined by the interaction potential $u_{\rm nn}(r)$.
Although  this has not been proved in the present study, we still are suggesting that
virial expansions for the  excluded volume pressure,
equation~(\ref{PexclVir}),
do not diverge in the range of temperatures  that are of interest for the parent fluid, i.e., the LJ-like HCAY fluid in this case, including temperatures that are below its critical point temperature but higher than its triple point temperature.
Some discrepancies between the virial EOS and computer experiment data that are observed in figure~\ref{Fig3} for densities $\rho\sigma^3 > 0.6$, should be attributed to the truncation of the virial EOS (\ref{PexclVir}) at the sixth virial coefficient.

The resulting augmented virial EOS, equation~(\ref{PaugVir}), has been tested for the LJ-like HCAY fluid in the wide density range and for temperature conditions that  were studied in the literature
so far, including those where the conventional virial EOS, equation~(\ref{Pvir}), exhibits difficulties~\cite{NareshSingh2009}.
Being rather accurate in the range of densities up to around $\rho\sigma^3 = 0.6$ at the supercritical temperatures,
the augmented virial EOS, equation~(\ref{PaugVir}), remains qualitatively correct at subcritical temperatures as well, showing no sign for divergence at the temperature as low as $T^*=1$.
Nevertheless, for making  final conclusion regarding the performance of the augmented virial EOS in the case of the LJ-like HCAY fluid, as well as for the potential application of this approach to investigate more complex and/or realistic class of fluid models, the evaluation of the higher order virial coefficients, namely,  $B_{\rm 7}^{\rm nn}(T)$, $B_{\rm 8}^{\rm nn}(T)$, $\dots$, $B_{\rm 10}^{\rm nn}(T)$ for the excluded volume
interaction energy $u_{\rm nn}(r)$, is highly desirable.

In general, the excluded volume pressure
could be obtained by means of equation~(\ref{pEXCL}), supposing that
molecular excluded volume $v_{\rm 0}$
as function of both the density and temperature is known.
Unfortunately, function $v_{\rm 0}(\rho, T)$  is not available in general case.
On other hand,  the pressure $p_{\rm 0}$ can be obtained from the knowledge of the forces that are responsible for excluded volume.
Namely, similarly to the case of the original van der Waals theory, when excluded volume pressure $p_{\rm 0}$ was identified with the pressure $p^{\rm hs}$ of the fluid system with a hard-sphere repulsion $u^{\rm hs}(r)$, the excluded volume pressure within the augmented van der Waals theory
can be obtained as the pressure $p^{\rm nn}$  of the fluid system with interaction potential $u_{\rm nn}(r)$.
These data can be extracted, for instance, from computer simulation experiment~\cite{Shukla2000} or
within the integral equation theories~\cite{blum,duh}. Such a route has been already explored~\cite{MelnykFPE2009,MelnykJSF2010}, resulting in the augmented van der Waals EOS for the LJ-like HCAY fluid.
The other possibility might be to utilize the excluded volume pressure
$p_{\rm 0}$ in the framework of the perturbed virial EOS approach~\cite{NezbedaAim1984,NezbedaSmith2004}.

\section*{Acknowledgements}

This work was supported by the Czech-Ukrainian Bilateral Cooperative Program.

\clearpage


\clearpage

\ukrainianpart

\title{Віріальні розклади та розширений метод Ван дер Ваальса: застосування до Леннард-Джонсівської моделі твердих сфер з притяганням Юкави}
\author{А. Трохимчук\refaddr{label1,label4}, Р. Мельник\refaddr{label1},
І. Незбеда\refaddr{label2,label3}}
\addresses{
\addr{label1} Iнститут фiзики конденсованих систем НАН України, вул. І. Свєнціцького, 1, 79011 Львiв, Україна
\addr{label4} Інститут прикладної математики та фундаментальних наук,
Національний університет ``Львівська політехніка'', 79013 Львiв, Україна
\addr{label2} Природничий факультет, Університет ім. Я.Е. Пуркинє, Усті над Лабем, 40096, Чеська Республіка
\addr{label3}  Лабораторія термодинаміки ім. Е. Хала,
Інститут фундаментальних хімічних процесів АН Чеської Республіки,
Прага-6, 16502, Чеська Республіка
}

\makeukrtitle

\begin{abstract}
\tolerance=3000%
Показано, що запропонований недавно [Melnyk et al., Fluid Phase Equilibr., 2009, \textbf{279}, 1] критерій розбиття потенціалу парної взаємодії на дві частини, одна з яких описує  як можна точніше виключений об'єм в системі,  приводить до виразів для віріальних коефіцієнтів, які суттєво покращують точність віріального рівняння стану в цілому та  для температур нижчих за критичну, зокрема. Як приклад, розглянуто застосування до Леннард-Джонсівської моделі твердих сфер з притяганням Юкави.
\keywords виключений об'єм, плин твердих сфер з притяганням Юкави, віріальне рівняння стану, другий віріальний коефіцієнт, рівняння Ван дер Ваальса

\end{abstract}

\end{document}